\documentstyle[aps,prl,epsf]{revtex}

\begin{document}
\title{Effect of Ring Exchange on Orbital Antiferromagnet}
\author{C. H. Chung, Hae-Young Kee and Yong Baek Kim}
\address{Department of Physics, University of Toronto, Ontario, Canada M5S 1A7}
\date{\today}
\maketitle

\begin{abstract}
We study the effect of four-particle ring exchange process
on orbital antiferromagnetic state that occurs in some
correlated electron systems in two dimensions.
The primary question is whether the ring exchange process
enhances or suppresses the orbital antiferromagnetic ordering.
Using the fact that the orbital antiferromagnetic state arises
in the large-$N$ limit of the $SU(N)$ generalization of the $t$-$J$ 
model, we consider the large-$N$ limit of the $t$-$J$-$K$ model 
where $K$ represents the four-particle ring exchange term. 
The phase diagrams in the
large-$N$ mean field theory are obtained for the half-filling
and finite hole concentrations at zero temperature.
It is found that the ring exchange in general favors dimerized
states or the inhomogeneous orbital antiferromagnetic state, and 
suppresses the homogeneous orbital antiferromagnetic state.
We compare our results with other related models of strongly
correlated systems with ring exchange processes.

\vspace{0.3cm}

\end{abstract}


\section{Introduction}
\label{sec:intro}
Ring exchange interactions are often present in strongly correlated
many-particle systems. In the Hubbard model of electrons with on-site
repulsive interaction $U$, the large $U$ limit leads to an antiferromagnetic
exchange interaction $J \sim t^2/U$ between electrons at nearby sites
where $t$ is the hopping amplitude of electrons. It was shown that the
next leading order term in the effective Hamiltonian is the four-particle
ring exchange interaction with the strength $K \sim t^3/U^2$ \cite{macdonald}.
The multi-particle ring exchange interactions have been studied
extensively in solid $^3$He \cite{he3}. It was also suggested
that the ring exchange interactions may play an important role in
understanding the `metal-insulator' transition in low density
two-dimensional electron systems \cite{klaus}.
More recently, ring exchange models have been revisited in the
context of the occurrence of novel quantum ground states in
strongly correlated electrons, bosons, and spin systems 
\cite{balents1,senthil,balents2,sandvik}.
In particular the possibility of fractionalized
phases where the elementary excitations carry fractions of quantum
numbers of electrons/spins has been discussed
\cite{balents1,senthil}. It was also suggested that
in some soft-core boson models with ring exchange interactions, there may
exist a novel phase which is neither a superfluid nor a conventional
insulator; the low energy excitations in this phase are described by
a continuous set of wavevectors with zero energy, which may be called
a Bose `surface' in analogy to the familiar Fermi surface of 
electrons \cite{balents2}.
On the other hand, a quantum Monte Carlo study of a quantum XY model with
a ring exchange interaction (which may correspond to a hard core version
of the boson model with a similar ring exchange considered in Ref.\cite{balents2})
did not find this novel phase, but found that new emerging ground
state is a striped bond-plaquette ordered state at intermediate values
of the ratio between the four-spin ring exchange and the nearest-neighbor
Heisenberg exchange couplings \cite{sandvik}.

Independent of these developments, there have been great interest in
finding novel broken symmetry phases in correlated electron systems.
Notable examples are the proposals of various time-reversal symmetry
broken phases for the explanation of the pseudogap phase of underdoped
cuprates.\cite{sudip,palee,varma}
One of these examples involve short-ranged (fluctuating) \cite{palee}
or long-ranged \cite{sudip} orbital antiferromagnetic order; the long-range 
ordered state is often called the staggered flux phase \cite{marston} or 
the d-density wave state \cite{sudip}.
This is a metallic state and characterized by alternating circulating
currents in plaquettes of the square lattice while the charge density
is uniform. Whether the short-range or the long-range ordered
orbital antiferromagnetism is responsible for the pseudogap in cuprates
has been a subject of recent debates \cite{sudip2,palee2}. 
Mean field theory studies of some microscopic models have shown that the 
orbital antiferromagnetic state appears in some part of the phase diagram 
\cite{marston,nayak}.
Some ladder models also seem to allow the orbital antiferromagnetic phase
\cite{schollwock}.

Ubiquitous presence of ring exchange interactions in strongly correlated
systems leads to the question of the stability of the orbital antiferromagnetic
state in the presence of ring exchange processes. This is a particularly
interesting question given that the large $U$ limit of the Hubbard model
possesses the ring exchange term \cite{macdonald}. 
Previous neutron scattering experiments
on an undoped cuprate La$_{2-x}$Sr$_x$CuO$_4$ suggests that the presence of 
ring exchange is necessary to explain the spin wave spectra\cite{aeppli}, 
while the effect of ring exchange interaction may become weaker at finite 
concentration of mobile holes. 
In this paper, we investigate the effect of ring exchange process
in a two-dimensional correlated electron system. In particular we would
like to understand the effect of ring exchange process on the orbital
antiferromagnetic state. A good starting point would be the large $N$ limit
of the $SU(N)$ generalization of the $t$-$J$ model where
the orbital antiferromagnetic state is already present as a possible
ground state. The ground states discovered in the large $N$ limit may be
better understood as possible `disordered' states of the zero temperature
magnetically ordered phases in the physical limit $N=2$ (such as spin
antiferromagnetism), as some parameter at $T=0$ changes or temperature is raised.
The stability of different phases in the large $N$ limit will be reflected
in those situations as well.

In this paper, we study the effect of the four-particle ring exchange interaction
in the $SU(N)$ generalization of the $t$-$J$-$K$ model in the large $N$ limit.
Here $K$ represents the strength of the four-particle ring exchange.
The phase diagrams in the large $N$ mean field theory are obtained at the
half-filling and at finite hole concentration for varying strength of $K$
and the Heisenberg exchange coupling $J$. We found that the ring exchange
interaction in general suppresses the homogeneous orbital antiferromagnetic 
state and favors bond-centered dimerized phases with broken translational 
symmetry and/or the inhomogeneous orbital antiferromagnetic state\cite{kee} 
depending on the value of $K$.

The rest of the paper is organized as follows. In section II, the $SU(N)$
generalization of the $t$-$J$-$K$ model and the corresponding large $N$
mean field theory are discussed. The phase diagrams for various values of
$J$ and $K$ at the half-filling and finite hole concentration are constructed
in section III. Summary of the results, discussions on possible relations
to other models with ring exchange, and conclusion can be found in section IV.

\section{The model} 
\label{sec:sun}

\begin{figure}
\epsfxsize = 4cm
\hspace{6cm}
\epsfbox{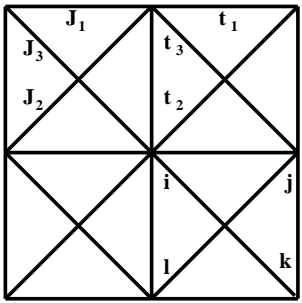}
\caption{Anisotropic triangular lattice with three types of the bonds.}
\label{lattice}
\end{figure}

We consider the $t$-$J$ model with four-particle exchange couplings on an 
anisotropic triangular lattice with nearest-neighbor and next-nearest-neighbor 
interactions:
\begin{eqnarray}
\label{HSU(2)}
H &=& -
t_1\sum_{<{\bf{ij}}>} c_{\bf{i}}^{\dagger\alpha} c_{\bf{j}\alpha} -
t_2\sum_{<<{\bf{jl}}>>}c_{\bf{j}}^{\dagger\alpha} c_{\bf{l}\alpha}-
t_3\sum_{<<{\bf{ik}}>>}c_{\bf{i}}^{\dagger\alpha} c_{\bf{k}\alpha}
\nonumber \\
&+&
J_1\sum_{<{\bf{ij}}>} (\vec{S}_{\bf{i}}\cdot \vec{S}_{\bf{j}} - 
\frac{1}{4} n_{\bf{i}} n_{\bf{j}}) + 
J_2\sum_{<<{\bf{jl}}>>} (\vec{S}_{\bf{j}}\cdot \vec{S}_{\bf{l}} - 
\frac{1}{4} n_{\bf{j}} n_{\bf{l}}) + 
J_3\sum_{<<{\bf{ik}}>>} (\vec{S}_{\bf{i}}\cdot \vec{S}_{\bf{k}} -
\frac{1}{4} n_{\bf{i}} n_{\bf{k}})
\nonumber \\ 
&+&
\sum_{<\bf{ijkl}>}[K 
(\vec{S}_{\bf{i}}\cdot \vec{S}_{\bf{j}})
(\vec{S}_{\bf{k}}\cdot \vec{S}_{\bf{l}}) +
(\vec{S}_{\bf{i}}\cdot \vec{S}_{\bf{l}})
(\vec{S}_{\bf{j}}\cdot \vec{S}_{\bf{k}}) -
\lambda
(\vec{S}_{\bf{i}}\cdot \vec{S}_{\bf{k}})
(\vec{S}_{\bf{j}}\cdot \vec{S}_{\bf{l}})]
, 
\end{eqnarray}
where $c_{\bf{i}\alpha}$ is the electron destruction operator of spin 
$\alpha = \uparrow, \downarrow$ on site $\bf{i}$, 
$\vec{S}_{\bf{i}} = \frac{1}{2} 
c_{\bf{i}}^{\dagger\alpha}{\vec \sigma}_{\alpha}^{\beta}c_{\bf{i} \beta}$ is 
the electron spin operator, and  
$n_{\bf{i}} = c_{\bf{i}}^{\dagger\alpha}c_{\bf{i}\alpha}$ (sum over
repeated indices are assumed throughout this paper) 
is the electron number operator.
Here $J_1$ and $t_1$ are the nearest-neighbor Heisenberg exchange 
coupling and hopping parameter while $J_2, J_3$ and $t_2, t_3$ are the
next-nearest-neighbor Heisenberg exchange coupling and hopping parameters 
respectively as shown in Fig.\ref{lattice}.
$K$ and $\lambda$ are the strength of the four-particle exchange couplings, 
and the sum in the four-particle exchange is over all non-overlapping 
plaquettes labelled clockwise by the sites $\bf{i}$, $\bf{j}$, $\bf{k}$ and $\bf{l}$ 
(see Fig.\ref{lattice}). Note that at $\lambda = K$, the four-particle 
exchange terms correspond to the conventional ring exchange process\cite{klaus}. 
To get further insight, we consider more general four-particle 
exchange interactions by treating $\lambda$ as an independent free 
parameter\cite{utsumi}. By doing so, we may at first investigate several simple 
limits where the model is easier to solve and the results in these limits can provide us 
useful guidance in studying the effect of various four-particle exchange 
couplings (including the ring exchange) on the orbital antiferromagnetic 
state. Then we will come back later to the case of more interesting and 
physical ring exchange coupling.

To make a progress, we apply a large-$N$ approach by generalizing the 
spin symmetry from $SU(2)$ to $SU(N)$ \cite{marston,subir91}. $N=2$ corresponds 
to the physical $SU(2)$ spin-$1/2$ system. The large-$N$ approach has been 
applied to various strongly correlated systems\cite{marston,subir91,chung}. 
One advantage of this approach is that the mean field solutions become 
exact in the limit of $N\to \infty$ and thus it is possible to systematically 
calculate corrections in powers of $1/N$. 

Let us briefly describe the $SU(N)$ generalization of the Hamiltonian 
in Eq.\ref{HSU(2)}. In the $SU(2)$ case, we can express the Hamiltonian
in terms of fermionic operators by use of the following identity\cite{marston}:
\begin{eqnarray}
\vec{S}_{\bf{i}}\cdot \vec{S}_{\bf{j}} 
&=& 
-\frac{1}{2}(c_{\bf{i}}^{\dagger\alpha}c_{\bf{j}\alpha})
(c_{\bf{j}}^{\dagger\beta}c_{\bf{i}\beta}) 
+ \frac{1}{2} n_{\bf{i}}
- \frac{1}{4} n_{\bf{i}}n_{\bf{j}},
\end{eqnarray}
The $SU(N)$ generalization of the $SU(2)$ Hamiltonian is obtained by simply
letting the spin index $\alpha$ run from 1 to $N$. We apply the slave-boson
method to impose the constraint in the Hilbert space. In the slave-boson 
representation, the spins are 
represented by fermions $f_{\alpha}$, $\alpha = 1\cdots N$ which 
transform under the fundamental representation of $SU(N)$; the holes are 
represented by spinless bosons $b_{\bf{i}}$ where 
$c_{\bf{i}\alpha} = f_{\bf{i}\alpha}b_{\bf{i}}^{\dagger}$. The local
constraint is given by\cite{vojta}:
\begin{equation}
f_{\bf{i}}^{\dagger\alpha}f_{\bf{i}\alpha} + 
b_{\bf{i}}^{\dagger}b_{\bf{i}} = \frac{N}{2}.
\end{equation}
In the $N\to \infty$ limit and at zero temperature, the average hole 
concentration $\delta$ can be obtained from \cite{vojta}
\begin{equation}
\frac{1}{N_s} \sum_{\bf{i}} <f_{\bf{i}}^{\dagger\alpha}f_{\bf{i}\alpha}> =
\frac{N}{2}(1 - \delta),
\end{equation}
where $N_s$ is the total number of sites in the system.
%
In this paper, we only consider 
the case with uniform charge distribution which excludes 
the possibility of site-centered charge density wave or stripe state 
that was found in some related models\cite{vojta}. Investigation of the 
possibility of charge stripe phases is an interesting and yet a separate 
topic which is beyond the scope of this paper. Based on this assumption, 
the hole concentration is given by $\delta$, and the electron density 
operator $n_{\bf{i}\alpha}$ is a constant ${\it{c}}$ number,
$n_{\bf{i}} = \frac{N}{2}(1 - \delta)$. Note that there is exactly one 
electron per site in the limit of the physical $SU(2)$ spin-1/2 case 
at the half-filling. 

The $SU(N)$ generalization of the Hamiltonian in Eq.\ref{HSU(2)} is therefore 
given by:
\begin{eqnarray}
\label{HSU(N)}
H_{SU(N)} &=& 
-t_{1}\delta\sum_{<\bf{ij}>}f_{\bf{i}}^{\dagger\alpha}f_{\bf{j}\alpha}
-t_{2}\delta\sum_{<<\bf{jl}>>}f_{\bf{j}}^{\dagger\alpha}f_{\bf{l}\alpha}
-t_{3}\delta\sum_{<<\bf{ik}>>}f_{\bf{i}}^{\dagger\alpha}f_{\bf{k}\alpha}
\nonumber \\
&-&
\frac{J_1^{e}}{N}
\sum_{<\bf{ij}>} 
|f_{\bf{i}}^{\dagger\alpha}f_{\bf{j}\alpha}|^2 - 
\frac{J_2^{e}}{N} 
\sum_{<<\bf{jl}>>} 
|f_{\bf{j}}^{\dagger\alpha}f_{\bf{l}\alpha}|^2 -
\frac{J_3^{e}}{N}
\sum_{<<\bf{ik}>>}
|f_{\bf{i}}^{\dagger\alpha}f_{\bf{k}\alpha}|^2 \nonumber \\
&+&
\sum_{<\bf{ijkl}>}[\frac{2K}{N^{3}}(
|f_{\bf{i}}^{\dagger\alpha}f_{\bf{j}\alpha}|^2
|f_{\bf{k}}^{\dagger\beta}f_{\bf{l}\beta}|^2
+
|f_{\bf{i}}^{\dagger\alpha}f_{\bf{l}\alpha}|^2
|f_{\bf{j}}^{\dagger\beta}f_{\bf{k}\beta}|^2)
- \frac{\lambda}{N^3}
|f_{\bf{i}}^{\dagger\alpha}f_{\bf{k}\alpha}|^2
|f_{\bf{j}}^{\dagger\beta}f_{\bf{l}\beta}|^2
], 
\end{eqnarray}
where the Heisenberg exchange couplings $J_{\bf{ij}}$ have been 
renormalized by the ring exchange interactions leading to the 
effective couplings: 
$J_{1}^{e} = 
J_1(1 - \frac{K}{2J_1}(1-\delta)^{2})$, $J_{2}^{e} = 
J_2(1 + \frac{\lambda}{4 J_2}(1-\delta)^{2})$, and $J_{3}^{e} =
J_3(1 + \frac{\lambda }{4 J_3}(1-\delta)^{2})$.
Note that we rescale both $J_{\bf{ij}}$ and $K$ by a factor of 
$\frac{2}{N}$ so that the Hamiltonian is of the order $N$.\cite{marston}. 
Also, we have dropped the constant terms involving $n_{\bf{i}}$ and 
$n_{\bf{i}}n_{\bf{j}}$ in the Hamiltonian.

The above Hamiltonian contains eight- and four-fermion interactions. 
We may decouple these interactions into quadratic interactions by 
applying the Hubbard-Stratonovich transformations. We then perform the 
functional integration over the Hubbard-Stratonovich fields in a 
saddle-point approximation\cite{marston}. The key point in the large-$N$ 
approach is that the saddle-point solution becomes exact in the 
$N\to \infty$ limit. The factorization of the spin-spin interactions 
can be done in two stages\cite{marston}. We first break up the eight-fermion 
terms by introducing real Hubbard-Stratonovich fields $\Phi_{\bf{ij}}$:
\begin{eqnarray}
\frac{2K}{N^3}
|f_{\bf{i}}^{\dagger\alpha}f_{\bf{j}\alpha}|^2
|f_{\bf{k}}^{\dagger\beta}f_{\bf{l}\beta}|^2 \rightarrow
N[-\frac{1}{2K}\Phi_{\bf{ij}}\Phi_{\bf{kl}} - 
               \frac{\Phi_{\bf{ij}}}{N^2} 
               |f_{\bf{k}}^{\dagger\beta}f_{\bf{l}\beta}|^2 - 
               \frac{\Phi_{\bf{kl}}}{N^2} 
               |f_{\bf{i}}^{\dagger\beta}f_{\bf{j}\beta}|^2]\nonumber \\
\frac{-2\lambda}{N^3}
|f_{\bf{i}}^{\dagger\alpha}f_{\bf{k}\alpha}|^2
|f_{\bf{j}}^{\dagger\beta}f_{\bf{l}\beta}|^2 \rightarrow
N[\frac{1}{2\lambda}\Phi_{\bf{ik}}\Phi_{\bf{jl}} - 
               \frac{\Phi_{\bf{ik}}}{N^2} 
               |f_{\bf{j}}^{\dagger\beta}f_{\bf{l}\beta}|^2 - 
               \frac{\Phi_{\bf{jl}}}{N^2} 
               |f_{\bf{i}}^{\dagger\beta}f_{\bf{k}\beta}|^2]
\end{eqnarray}
At the saddle-point $\Phi_{\bf{ij}}$ fields take the following values:
\begin{eqnarray}
\Phi_{\bf{ij}} &=& -\frac{2K}{N^2}
<|f_{\bf{i}}^{\dagger\alpha}f_{\bf{j}\alpha}|^2>\nonumber \\
\Phi_{\bf{ik}} &=& \frac{2\lambda}{N^2}
<|f_{\bf{i}}^{\dagger\alpha}f_{\bf{k}\alpha}|^2>,
\end{eqnarray}  
 where $\Phi_{\bf{ij}}$ and $\Phi_{\bf{ik}}$ represent 
the fields along the nearest-neighbor and next-nearest-neighbor
sites respectively.

After the first Hubbard-Stratonovich transformation, the Hamiltonian
now still contains four-fermion interactions. We can decouple these 
terms by introducing the complex Hubbard-Stratonovich fields $\chi_{\bf{ij}}$:
\begin{eqnarray}
-J_{1}^{e} (1+ \frac{2\Phi_{\bf{kl}}}{J_{1}^{e}})  
|f_{\bf{i}}^{\dagger\alpha}f_{\bf{j}\alpha}|^2 
&\rightarrow&
(1+ \frac{2\Phi_{\bf{kl}}}{J_{1}^{e}})  
(\frac{|\chi_{\bf{ij}}|^2}{J_{1}^{e}} - 
\frac{\chi_{\bf{ij}}^{\ast}}{N}
f_{\bf{i}}^{\dagger\alpha}f_{\bf{j}\alpha} + H.c.),\nonumber \\  
-J_{2}^{e} (1+ \frac{\Phi_{\bf{ik}}}{J_{2}^{e}})  
|f_{\bf{j}}^{\dagger\alpha}f_{\bf{l}\alpha}|^2 
&\rightarrow&
(1+ \frac{\Phi_{\bf{ik}}}{J_{2}^{e}})  
(\frac{|\chi_{\bf{jl}}|^2}{J_{2}^{e}} - 
\frac{\chi_{\bf{jl}}^{\ast}}{N}
f_{\bf{j}}^{\dagger\alpha}f_{\bf{l}\alpha} + H.c.),\nonumber \\
-J_{3}^{e} (1+ \frac{\Phi_{\bf{jl}}}{J_{3}^{e}})  
|f_{\bf{i}}^{\dagger\alpha}f_{\bf{k}\alpha}|^2 
&\rightarrow&
(1+ \frac{\Phi_{\bf{jl}}}{J_{3}^{e}})
(\frac{|\chi_{\bf{ik}}|^2}{J_{3}^{e}} -
\frac{\chi_{\bf{ik}}^{\ast}}{N}
f_{\bf{i}}^{\dagger\alpha}f_{\bf{k}\alpha} + H.c.).  
\end{eqnarray}
where $\chi_{\bf{ij}}$ fields have the following saddle-point
values\cite{marston}:
\begin{equation}
\chi_{\bf{ij}} = \frac{J_{\bf{ij}}^{e}}{N}
<f_{\bf{i}}^{\dagger\alpha}f_{\bf{j}\alpha}>.
\end{equation} 

After two Hubbard-Stratonovich transformations, the Hamiltonian is now 
quadratic in fermion operators. Note that at the saddle-point 
$\Phi_{\bf{ij}}$ and $\chi_{\bf{ij}}$ fields have the following relations:
\begin{eqnarray}
\Phi_{\bf{ij}} = -\frac{2K}{(J_{\bf{ij}}^{e})^2}|\chi_{\bf{ij}}|^2\nonumber \\
\Phi_{\bf{ik}} = \frac{2\lambda}{(J_{\bf{ik}}^{e})^2}|\chi_{\bf{ik}}|^2,
\end{eqnarray}
where ${\bf{ij}}$ and ${\bf{ik}}$ again are the nearest-neighbor and 
next-nearest-neighbor sites respectively. Therefore, we 
can express the mean-filed Hamiltonian purely in terms of 
$\chi$-fields. After integrating out the fermions, the 
resulting free energy is given by\cite{marston}:
\begin{eqnarray}
\frac{E_{MF}}{N} &=& \sum_{<\bf{ijkl}>} [
-\frac{6K}{(J_{1}^{e})^4}
(|\chi_{\bf{ij}}|^2 |\chi_{\bf{kl}}|^2 +
 |\chi_{\bf{il}}|^2 |\chi_{\bf{jk}}|^2 ) +
\frac{6K}{(J_{2}^{e})^2(J_3^{e})^2}
|\chi_{\bf{ik}}|^2 |\chi_{\bf{jl}}|^2]\nonumber \\
&+&
\sum_{<\bf{ij}>}\frac{|\chi_{\bf{ij}}|^2}{J_{1}^{e}} +
\sum_{<<\bf{jl}>>}\frac{|\chi_{\bf{jl}}|^2}{J_{2}^{e}} +
\sum_{<<\bf{ik}>>}\frac{|\chi_{\bf{ik}}|^2}{J_{3}^{e}}\nonumber \\
&-&
\frac{1}{\beta}\sum_{\bf k} \ln \{1 + e^{[-\beta(\omega_{\bf k} - \mu)]}\},
\end{eqnarray}
where $\mu$ is the chemical potential which fixes the number of the fermions, 
$\beta = \frac{1}{k_{\bf{B}}T}$ is the inverse of the temperature, 
$\omega_{\bf k}$ are the eigenvalues of the following Hamiltonian $H_1$: 
\begin{eqnarray}
\frac{H_{1}}{N} &=& 
-t_{1}\delta\sum_{<\bf{ij}>}f_{\bf{i}}^{\dagger\alpha}f_{\bf{j}\alpha}
-t_{2}\delta\sum_{<<\bf{jl}>>}f_{\bf{j}}^{\dagger\alpha}f_{\bf{l}\alpha}
-t_{3}\delta\sum_{<<\bf{ik}>>}f_{\bf{i}}^{\dagger\alpha}f_{\bf{k}\alpha}
\nonumber \\
&+&
\sum_{<\bf{ijkl}>}[
(1-\frac{4K|\chi_{\bf{kl}}|^2}{(J_{1}^{e})^3})  
(- \frac{\chi_{\bf{ij}}^{\ast}}{N}
f_{\bf{i}}^{\dagger\alpha}f_{\bf{j}\alpha} + H.c.)\nonumber \\ 
&+&
(1- \frac{4K|\chi_{\bf{il}}|^2}{(J_{1}^{e})^3})  
(- \frac{\chi_{\bf{jk}}^{\ast}}{N}
f_{\bf{j}}^{\dagger\alpha}f_{\bf{k}\alpha} + H.c.)\nonumber \\ 
&+&
(1+ \frac{2\lambda |\chi_{\bf{ik}}|^2}{J_{2}^{e}(J_3^{e})^2})  
(- \frac{\chi_{\bf{jl}}^{\ast}}{N}
f_{\bf{j}}^{\dagger\alpha}f_{\bf{l}\alpha} + H.c.)\nonumber \\
&+&
(1+ \frac{2\lambda |\chi_{\bf{jl}}|^2}{J_{3}^{e}(J_2^{e})^2})  
(- \frac{\chi_{\bf{ik}}^{\ast}}{N}
f_{\bf{i}}^{\dagger\alpha}f_{\bf{k}\alpha} + H.c.)
].
\end{eqnarray}

\begin{figure}
\epsfxsize = 5cm
\hspace{6cm}
\epsfbox{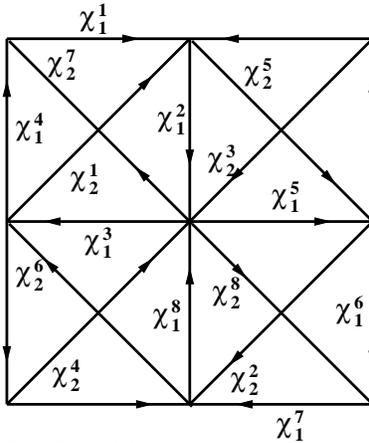}
\caption{The 2 $\times$ 2-site unit cell of the $SU(N)$ $t$-$J$ model on an 
anisotropic triangular lattice. Arrows indicate the orientation of the 
complex-valued $\chi_{\bf{ij}}$ fields.}
\label{chi}
\end{figure}

Note that at the half-filling, the mean-field Hamiltonian is invariant under  
local U(1) gauge transformations as long as the $\chi$-fields 
transform as gauge fields\cite{marston}: 
$\chi_{\bf ij}(\tau) \to e^{i[\theta_{\bf i}(\tau) - 
\theta_{\bf j}(\tau)]} \chi_{\bf ij}(\tau)$. Away from half-filling, this 
local U(1) gauge symmetry is broken down to only the global U(1) gauge symmetry 
reflecting the conservation of total charge. The ground state of the 
system is found by solving the saddle-point equations, 
$\frac{\partial{E_{MF}}}{\partial{\chi_{\bf{ij}}}} = 0,$ 
for $\chi_{\bf{ij}}$-fields. Extrema of $E_{MF}$ are found numerically 
with the simplex-annealing method\cite{NRC}. To make progress in 
solving the model, we make the assumption that spontaneous symmetry 
breaking, if it occurs, does not lead to a unit cell larger than 4 sites. 
Our choice of unit cell is shown in Fig.\ref{chi}. The $2 \times 2$ unit 
cell requires 16 different complex-valued $\chi$-fields ($8$ $\chi_1$-fields 
on the square and $8$ $\chi_2$-fields along the diagonals). We work with a 
lattice of $40 \times 40 $ sites and check that this is sufficiently large to 
accurately represent the thermodynamic limit. We check all the saddle-point 
solutions and make sure that the ground state is the one with the lowest 
free energy.  

\section{Zero temperature phase diagram}
\label{sec:phase}

Possible ground states of the system may be classified into several different
phases which are characterized by the symmetry of the $\chi_{\bf{ij}}$-
fields at the saddle-point. In the half-filled ($\delta = 0$) case, it is 
important to classify the phases in a gauge-invariant fashion 
because there are many gauge-equivalent saddle-points. There are two 
important gauge-invariant quantities\cite{chung}: 

\noindent (i) The squared amplitude $|\chi_{\bf ij}|^2$ which is proportional to the 
spin-spin correlation function $\langle \vec{S}_{\bf i} \cdot \vec{S}_{\bf j} \rangle$.
Modulations in $|\chi|$ signal the presence of a bond-centered dimerization.

\noindent (ii) The plaquette operator $\Pi \equiv \chi_{12} \chi_{23} \chi_{34}
\chi_{41}$, where  1, 2, 3, and 4 are sites on the corners of a unit
plaquette. By identifying the phase of $\chi$ as a spatial gauge field
it is clear that the plaquette operator is gauge-invariant, and its phase 
measures the amount of magnetic flux penetrating the plaquette
\cite{marston}. Different saddle points are therefore gauge equivalent if 
the plaquette operator has the same expectation value, even though the 
$\chi$-fields may be different. At half-filling the flux always equals 
to $0$ or $\pi$ (mod $2 \pi$), so one can always choose a gauge such that
all the $\chi$-fields are purely real \cite{marston}.  

Away from the $\delta = 0$ limit, the saddle-point 
solutions are further classified in terms of whether or not they break 
time-reversal symmetry ($\hat{T}$) where the plaquette operator has 
a nontrivial phase other than $0$ or $\pi$. The $\hat{T}$-breaking state
will generate real orbital currents going through the plaquettes in an 
alternating fashion \cite{marston,chung}. Finally, as there are eight 
independent parameters ($t_1$, $t_2$, $t_3$, $J_1$, $J_2$, $J_3$, $K$, 
and $\lambda$) the phase diagram lives in a seven-dimensional space of 
their dimensionless ratios. Since we are interested in how the four-particle 
exchange couplings may affect the orbital antiferromagnetic state, we 
reduce the parameter space to a more manageable two-dimensional section by fixing 
the ratios of $J_1^e/J_2^e$, $J_1^e/J_3^e$, $t_1/t_2$, $t_1/t_3$, and $K/\lambda$
such that the orbital antiferromagnetic state appears as one of the ground states
of the system. Then by varying $J_1^e/t_1$ (or $J_2^e/t_2$) and 
$K/J_1^e$ (or $\lambda/J_2^e$) we explore the phase diagram in a two-dimensional 
space. We summarize the phases which appear in the phase diagram in the following 
subsections.

\subsection{Half-filling ($\delta = 0$)}

We find that in this limit the ground states are always 
dimerized in one way or another with spatial modulations in spin-spin 
correlations. Although the staggered flux phase (that becomes the orbital 
antiferromagnetic state at non-zero hole concentration) is a saddle-point 
solution, its free energy is always higher than the 
dimerized ground states. To see this more clearly, we plot the free energy 
as a function of the four-particle exchange parameter $K$ in 
the square lattice limit ($J_2^e = J_3^e = 0$) with $\lambda = 0$ 
(see Fig.\ref{free1}). We compare the free 
energies of the saddle-point solutions which are competing with each other 
to be the true ground state. In this limit, we find the following 
competing phases:

{\bf{Dimer phase}}

\begin{figure}
\epsfxsize = 3cm
\hspace{6cm}
\epsfbox{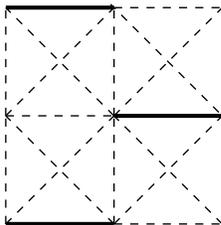}
\caption{The dimer phase with decoupled dimerized spin chains. 
Dark solid links indicate
nonzero spin-spin correlations and dash links have zero spin-spin 
correlations.}
\label{dimersq0}
\end{figure}

This is a fully dimerized phase which exhibits spin-Peierls 
order: $|\chi_1^1| = |\chi_1^5| \not= 0$, and all other $\chi$-fields are zero;
See Fig.\ref{dimersq0} for a sketch\cite{chung}. It breaks the translational 
symmetry and have spatial modulations in spin-spin correlations. In fact, 
the system breaks up into decoupled dimerized spin chains. It is an insulating 
phase as there is a large gap in the energy spectrum at the Fermi energy. 

{\bf{Box phase}}

\begin{figure}
\epsfxsize = 3cm
\hspace{6cm}
\epsfbox{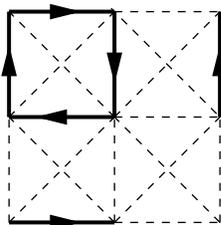}
\caption{Box phase with $0$ or $\pi$ flux around the plaquette.
Dark solid links indicate
nonzero spin-spin correlations and dash links have zero spin-spin 
correlations. The arrows indicate 
the complex-valued $\chi$-fields.}
\label{boxsq0}
\end{figure}

This insulating phase breaks the translational symmetry and 
consists of isolated plaquettes with enhanced spin-spin 
correlations\cite{chung,dombre,ssmodel}. The $\chi_1$-fields are either 0 or 
complex with $|\chi_1^1| = |\chi_1^2| = |\chi_1^3| = |\chi_1^4| \not= 0$, 
and $|\chi_1^5| = |\chi_1^6| = |\chi_1^7| =|\chi_1^8| = 0$. All $\chi_2$ 
fields are zero. See Fig.\ref{boxsq0} for a sketch. This phase is similar 
to the fully dimerized phase as the magnitudes of the $\chi_1$-fields
modulate on the lattice. The box phase does not break the time-reversal 
symmetry as the phase of the plaquette product $\chi_1^1 \chi_1^2 \chi_1^3 \chi_1^4$ 
is either 0 or $\pi$. There are no real orbital currents circulating since we can 
always choose a gauge to make $\chi$-fields real.

{\bf{Staggered Flux Phase}}

\begin{figure}
\epsfxsize = 3cm
\hspace{6cm}
\epsfbox{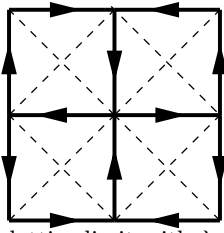}
\caption{Staggered flux phase in the square lattice limit with $\lambda = 0$. 
The dark links on the square indicate strong spin-spin correlations and 
the dash links along the diagonals have zero spin-spin correlation.
The arrows indicate the orientation of the complex-valued $\chi$-fields.
At half-filling, there is no real circulating orbital currents since the 
flux around the plaquette is either $0$ or $\pi$. Away from half-filling, 
however, the flux around the plaquette is neither $0$ nor $\pi$. There are 
therefore real orbital currents in this case circulating in an antiferromagnetic 
fashion as indicated by the arrows; thus it is an orbital antiferromagnet.}
\label{fluxsq0}
\end{figure}

\begin{figure}
\epsfxsize = 9cm
\hspace{3cm}
\epsfbox{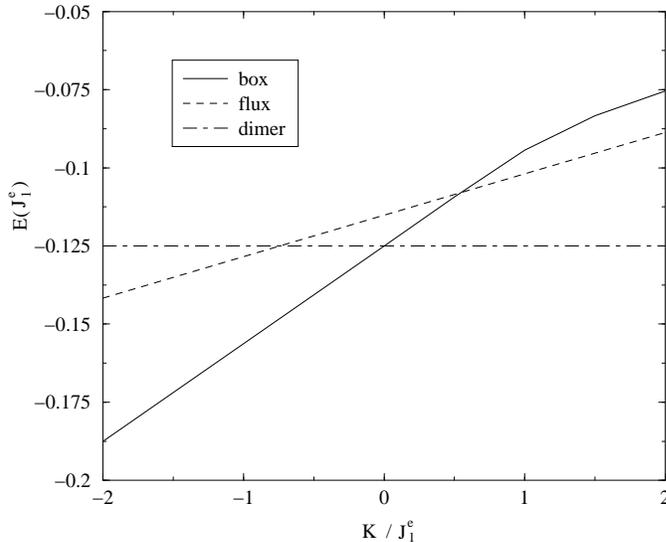}
\caption{Free energy of the half-filled $t-J$ model on the square lattice
as a function of $K/J_1^e$ for three competing saddle-point solutions.
Note that the dimer and box phase are degenerate ground states at $K = 0$.}
\label{free1}
\end{figure}

All $\chi_1$-fields are equal, with an imaginary component in general, 
and all $\chi_2$-fields along the diagonals are zero; 
see Fig.\ref{fluxsq0} for a sketch. Like the box phase, the staggered flux 
phase at the half-filling does not break $\hat{T}$-symmetry as the phase 
of the plaquette operator is either 0 or $\pi$. We can always make a gauge 
transformation such that all $\chi_1$-fields are real. There are therefore 
no real orbital currents circulating around the plaquettes. This phase is 
semi-metallic as the density of states is small, and in fact vanishes 
linearly at the Fermi energy. The gauge fluctuations at sufficiently 
small-$N$ are expected to drive the staggered flux phase into a 
N\'eel-ordered state\cite{gauge}.  

We plot the free energy as a function of $K/J_1^e$ as shown in 
Fig.\ref{free1}. We can see that the box and the dimer phases are
the ground states for $K < 0$ and $K > 0$ respectively. The 
staggered flux phase is not the ground state in both cases.
We have also checked the free energy as a function of $\lambda/J_2^e$ 
in the other extreme limit where $J_1^e = 0$ and $K = 0$ 
(two inter-penetrating square lattice limit). We also find that 
the staggered flux phase is not the ground state for both $\lambda > 0$ 
and $\lambda < 0$.

From our results at the half-filling, it is clear that 
the four-particle exchange actually favors the dimerized phases over
the staggered flux phase. The dimerized state which breaks translational symmetry
is in fact a bond-centered charge density wave state\cite{marston}. Recent 
Monte Carlo study of the quantum XY model with ring exchange indicates 
plaquette (stripe) order which is also a bond-centered charge density wave 
state as the ring exchange coupling exceeds the critical value\cite{sandvik}.
In fact, the box phase in our model in the square lattice
limit with $K < 0$ looks very similar to the plaquette phase in the quantum 
XY model mentioned above. Although there are differences between the XY model and 
our $t$-$J$ model, these two models share the same structure of the 
spin-spin interactions and have similar ground states. Note that the 
four-particle exchange and the ring exchange couplings share 
the similar form of spin-spin interactions. We therefore expect that even 
though the staggered flux phase is not a ground state in our model at 
half-filling, our result at half-filling can still give us an important
clue about the effects of the ring exchange couplings on the system at 
finite doping where the staggered flux phase does appear to be a ground state. 
Now we proceed to solve our model at finite doping.

\subsection{Nonzero doping ($\delta = 0.05$)}

Unlike the half-filled case where the staggered flux phase is never 
a ground state, at nonzero doping the hoping terms are expected to 
stabilize the staggered flux phase\cite{marston}. We find the staggered flux 
phase is indeed the ground state of our system over a range of 
$J_1^e/t_1$ ratio. From our results at the half-filling, however, we know 
that the four-particle exchange favors dimerization over the 
staggered flux phase. We find it is also the case at finite doping. 
There is therefore the competition between the hopping terms and 
the four-particle exchange couplings which results in 
an interesting phase diagram.   

Again, we first work on the phase diagrams of two simple limits 
where the staggered flux phase appears as a stable ground state. 
We then move on to the more interesting ring exchange limit where all the 
couplings are nonzero. Here, we fix the doping concentration $\delta$ 
to be $0.05$.

\subsubsection{Square lattice limit with $\delta = 0.05$}

As shown in Fig.\ref{sq0}, we find the following phases in our phase 
diagram:

{\bf{Dimer phase}}

\begin{figure}
\epsfxsize = 3cm
\hspace{6cm}
\epsfbox{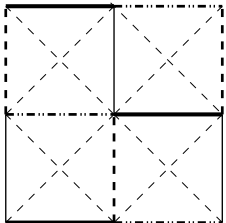}
\caption{The one-dimensional partially dimerized phase at $\delta = 0.05$. 
Dark solid links indicate strong spin-spin correlations, light solid and 
dark dash links indicate weak spin-spin correlations. The $\chi_2$-fields 
on the diagonals are zero.}
\label{dimersq5}
\end{figure}

This dimer phase breaks 
translational symmetry and it exists at large $J_1^e/t_1$ and positive 
$K/J_1^e$ ratio. This phase is a partially dimerized phase
 as $|\chi_1^1| = |\chi_1^5| > |\chi_1^2| = |\chi_1^6| > 
|\chi_1^3| = |\chi_1^7| >|\chi_1^4| = |\chi_1^8|$. All $\chi_2$-fields 
are zero. See Fig.\ref{dimersq5} for a sketch.

{\bf{Box phase}}

\begin{figure}
\epsfxsize = 3cm
\hspace{6cm}
\epsfbox{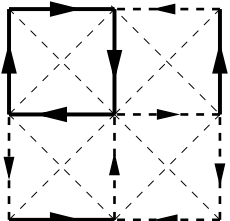}
\caption{Box phase on the square lattice at $\delta = 0.05$.
The flux around the plaquette is neither $0$ nor $\pi$.
Dark solid links indicate larger magnitudes of the $\chi_1$-fields or 
stronger spin-spin correlations and dark dash links indicate smaller 
magnitudes of the $\chi_1$-fields or weaker spin-spin correlations. 
The arrows indicate the complex-valued $\chi_1$-fields and also the 
directions of the orbital currents. The $\chi_2$-fields along the 
diagonals are zero. We have checked that the orbital currents 
are conserved at each site.}
\label{boxsq5}
\end{figure}

This box phase next to the dimer phase also breaks the translational 
symmetry and exists at large $J_{1}^{e}/t_1$ ratio and at negative
or small positive $K/J_{1}^e$ ratio. This phase is very similar 
to the box phase at half-filling except now it is partially dimerized.
All $\chi_1$-fields are complex and 
$|\chi_1^1| = |\chi_1^2| = |\chi_1^3| = |\chi_1^4| > |\chi_1^5| = 
|\chi_1^6| = |\chi_1^7| =|\chi_1^8|$. The imaginary parts of $\chi_1$-fields 
have the similar relation: $Im(\chi_1^1) = Im(\chi_1^2) = 
Im(\chi_1^3) = Im(\chi_1^4) > Im(\chi_1^5) =
Im(\chi_1^6) = Im(\chi_1^7) = Im(\chi_1^8)$. All $\chi_2$-fields 
along the diagonals are zero.
This box phase at finite doping breaks the time-reversal symmetry as
the phase of the plaquette product is neither 0 nor $\pi$ in general. 
There are {\it real} orbital currents circulating around the plaquette. 
This corresponds to an inhomogeneous orbital antiferromagnetic state. 
See Fig.\ref{boxsq5} for a sketch.

{\bf{Staggered flux phase (Homogeneous orbital antiferromagnetic state)}}

This phase exists at intermediate range of $J_1^e/t_1$ ratio.
It is very similar to the staggered flux phase at half-filling
except that now it breaks $\hat{T}$-symmetry as the phase of the plaquette 
operator is neither 0 nor $\pi$; see Fig.\ref{fluxsq0} for a sketch.
There are therefore real orbital currents circulating around the plaquette.
Note that unlike the box phase, the $\chi$-fields in the staggered 
flux phase have the same magnitude.

{\bf{Uniform phase}}

\begin{figure}
\epsfxsize = 3cm
\hspace{6cm}
\epsfbox{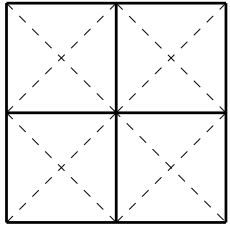}
\caption{Uniform phase in the square lattice limit. This phase does not
have broken symmetries and it is a Fermi liquid.}
\label{uniformsq}
\end{figure}

\begin{figure}
\epsfxsize = 9cm
\hspace{3cm}
\epsfbox{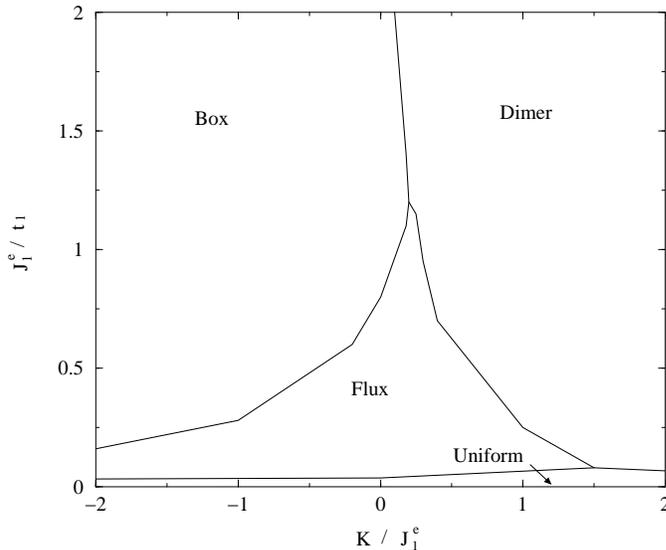}
\caption{Zero temperature phase diagram of 
the $SU(N)$ $t$-$J$ model in the square lattice
limit as a function of $J_1^e/t_1$ and $K/J_1^e$.}
\label{sq0}
\end{figure}

This phase exists at small $J_1^e/t_1$ ratio where all $\chi_1$-fields 
become negative real numbers and $\chi_1^1 = \chi_1^2 = \cdots = \chi_1^8$, 
and all $\chi_2$-fields are zero. See Fig.\ref{uniformsq} for a sketch. 
This phase preserves $\hat{T}$-symmetry and since all $\chi$-fields are real, 
they simply renormalize the hopping parameter $t_1$. The uniform phase is 
therefore a metallic Fermi liquid. Spin-spin correlations in the uniform 
phase decay with an inverse power law of the separation \cite{chung}.  

As shown in Fig.\ref{sq0}, the general features of the phase diagram 
suggest that the region of the staggered flux phase (homogeneous orbital 
antiferromagnetic state) becomes narrower and that
of the dimerized phase and/or the box phase (inhomogeneous orbital 
antiferromagnetic state) becomes wider as we increase both the positive and 
negative $K/J_1^e$ ratios. This is consistent with what we find at 
the half-filled case where the four-particle exchange couplings enhance 
dimerization and suppress the staggered flux phase. Also, for a fixed 
$J_1^e/t_1$ ratio, the ground state may change from the box phase to the 
staggered flux phase and finally to the dimer phase as the ratio of 
$K/J_1^e$ is changed from negative to positive values.

\subsubsection{Two inter-penetrating 
square lattice limit with $\delta = 0.05$}

In this limit, the ground state phases are classified as follows:

{\bf{1D dimer phase}}

\begin{figure}
\epsfxsize = 3cm
\hspace{6cm}
\epsfbox{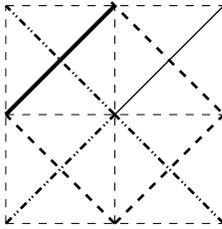}
\caption{The `one-dimensional' partially dimerized phase along 
the diagonals. Solid links indicate
strong spin-spin correlations and dark dash links indicate weak 
spin-spin correlations. The $\chi_1$-fields on the square are zero.}
\label{1ddimer5}
\end{figure}

This phase is a `one-dimensional' partially dimerized phase which breaks 
up the system into nearly decoupled dimerized spin chains 
along one of the two diagonal directions: 
$|\chi_2^1| > |\chi_2^3| > |\chi_2^4| = |\chi_2^7| = |\chi_2^8|
> |\chi_2^2| = |\chi_2^5| = |\chi_2^6|$ (see Fig.\ref{1ddimer5}). 
This phase is the ground state for $\lambda < 0$. Note that at half-filling, 
this phase is completely dimerized: $|\chi_2^1|=|\chi_2^3|\not=0$, and 
all other $\chi$-fields are zero.

{\bf{2D dimer phase}}

\begin{figure}
\epsfxsize = 3cm
\hspace{6cm}
\epsfbox{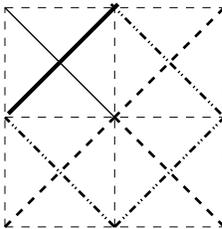}
\caption{The `two-dimensional' partially dimerized phase along the diagonals. 
Solid links indicate strong spin-spin correlations and dark dash links 
have weak spin-spin correlations. The $\chi_1$-fields on the square are zero.}
\label{2ddimer5}
\end{figure}

This phase is a `two-dimensional' partially dimerized phase which breaks 
up the system into nearly decoupled dimerized spin chains 
along both of the diagonal directions: : 
$|\chi_2^1| > |\chi_2^7| > |\chi_2^3| = |\chi_2^4| = |\chi_2^8|
> |\chi_2^2| = |\chi_2^5| = |\chi_2^6|$ (see Fig.\ref{2ddimer5}). 
This phase is the ground state for $\lambda > 0$. Note that at 
half-filling, this phase is completely dimerized: 
$|\chi_2^1|=|\chi_2^7|\not=0$, and all other $\chi$-fields are zero.

{\bf{Staggered flux phase (Homogeneous orbital antiferromagnetic phase)}}

\begin{figure}
\epsfxsize = 3cm
\hspace{6cm}
\epsfbox{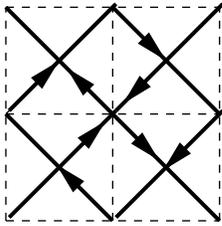}
\caption{Staggered flux phase in the two inter-penetrating square 
lattice limit. The dark links along the diagonals indicate strong spin-spin 
correlations and the dash links on the square have zero spin-spin correlation.
The arrows indicate the orientation of the complex-valued $\chi$-fields.
The lattice can be also viewed in this limit as if there are two inter-penetrating 
square lattices rotated by $45^\circ$. At half-filling, there is no real 
circulating orbital currents since the flux around the plaquette of these 
rotated square lattices is either $0$ or $\pi$. Away from half-filling, however, 
the flux is neither $0$ nor $\pi$. Therefore, there are real orbital currents
circulating in an antiferromagnetic fashion as indicated by the arrows.}
\label{fluxdbsq0}
\end{figure}

The staggered flux phase exists at intermediate values of 
$J_2^e/t_2$.
All $\chi_2$-fields are equal, with an imaginary component in general, 
and all $\chi_1$-fields on the square are zero (see Fig.\ref{fluxdbsq0}). 
This phase is very similar to the staggered flux phase on 
the square lattice except that now the flux goes through the plaquettes of 
two square lattices (rotated by $45^\circ$) inter-penetrating to each other.
Like staggered flux phase on the square lattice at nonzero doping, 
this phase also breaks $\hat{T}$-symmetry and there are real orbital 
currents running through the two inter-penetrating plaquettes as the 
phase of the plaquette operator is neither 0 nor $\pi$.

{\bf{Uniform phase}}

\begin{figure}
\epsfxsize = 3cm
\hspace{6cm}
\epsfbox{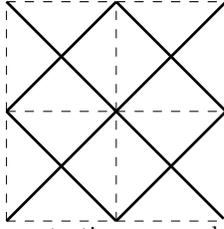}
\caption{Uniform phase in the two inter-penetrating square lattice limit at 
$\delta = 0.05$. This phase does not have broken symmetries and it is a 
Fermi liquid.}
\label{uniformdbsq}
\end{figure}

This is the ground state at small $J_2^e/t_2$ ratio. 
This metallic phase is similar to the uniform phase in the square lattice 
limit except that now $\chi_2^1 = \cdots = \chi_2^8\not= 0$ and all 
$\chi_1$-fields are zero (see Fig.\ref{uniformdbsq}). 

\begin{figure}
\epsfxsize = 9cm
\hspace{3cm}
\epsfbox{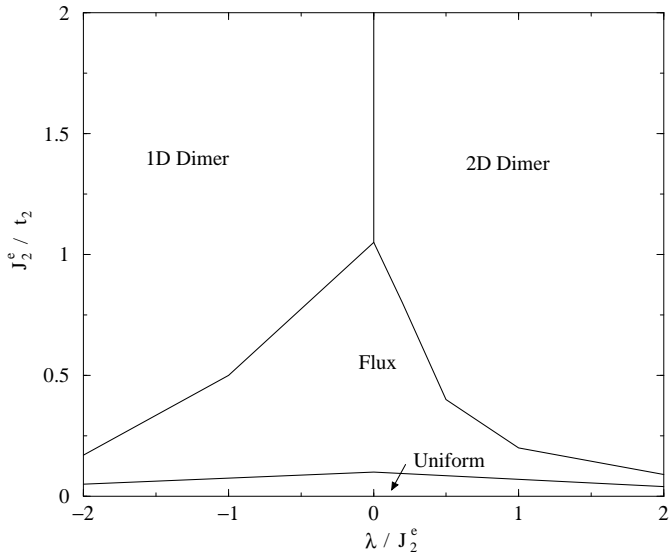}
\caption{Zero temperature phase diagram of the $SU(N)$ $t$-$J$ model 
in the limit of $J_1^e=t_1=K=0$, and $J_2^e=J_3^e$, $t_2=t_3$. 
The doping concentration $\delta = 0.05$.}
\label{dbsq}
\end{figure}

We again plot the phase diagram as a function of $J_2^e/t_2$ and 
$\lambda/J_2^e$ as shown in Fig.\ref{dbsq}. The phase diagram suggests 
that the region of the staggered flux phase becomes narrower and that 
of the dimerized phases becomes wider as both the positive and negative 
$\lambda/J_2^e$ ratios are increased. We reach the same conclusion 
as in the square lattice limit; the four-particle exchange couplings 
enhance dimerization and suppress the staggered flux phase (homogeneous 
orbital antiferromagnetic state). Also, 
for a fixed $J_1^e/t_1$ ratio, there are quantum phase transitions between 
the 1D dimer phase and the staggered flux phase as well as between 
staggered flux phase and the 2D dimer phase as the ratio of 
$\lambda/J_2^e$ is varied.

In addition to the above two extreme limits, we also look at
the more interesting ring exchange limit where $K = \lambda$ and all 
$J_{\bf{ij}}$ and $t_{\bf{ij}}$ are nonzero. We explore a wide range of 
the parameter space and find that the staggered flux phase only exists 
in the region close to the above two extreme limits. It is expected 
because the staggered flux phase is suppressed by geometrical frustration 
effect in other regions of the phase diagram \cite{chung}. 
We therefore restrict our attention to the parameter space close to these 
two limits where the staggered flux phase exists as a stable ground state.

\begin{figure}
\epsfxsize = 9cm
\hspace{3cm}
\epsfbox{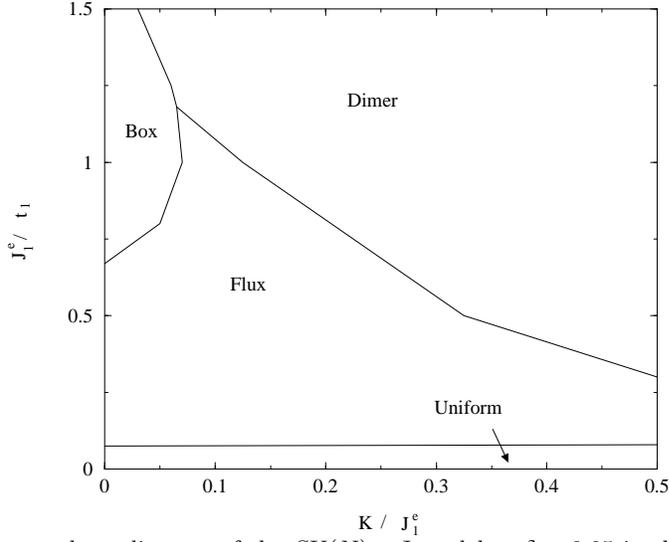}
\caption{Zero temperature phase diagram of the $SU(N)$ $t$-$J$ model 
at $\delta = 0.05$ in the ring exchange ($K=\lambda$) limit. 
Here, $J_2^e/J_1^e=J_3^e/J_1^e=0.5$ and $t_2/t_1=t_3/t_1=0.1$.
All the phases here are very similar to the corresponding phases
in the square lattice limit except that now the $\chi_2$-fields 
along the diagonals take small nonzero real values due to 
frustration.}
\label{sq05}
\end{figure}

\begin{figure}
\epsfxsize = 9cm
\hspace{3cm}
\epsfbox{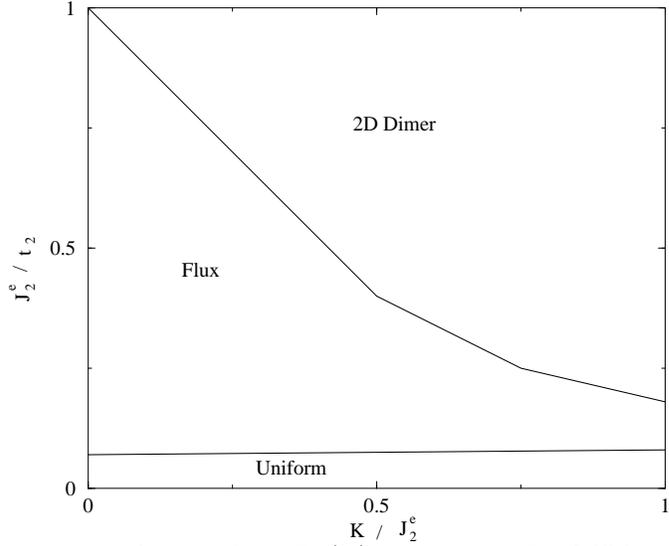}
\caption{Zero temperature phase diagram of the $SU(N)$ $t$-$J$ model 
at $\delta = 0.05$ in the ring exchange ($K=\lambda$) limit. 
Here, $J_1^e/J_2^e=J_1^e/J_3^e=0.25$ and $t_1/t_2=t_1/t_3=0.5$. 
All the phases here are very similar to the corresponding phases
in the limit of two inter-penetrating square lattice except 
that now the $\chi_1$-fields on the square take small nonzero 
real values due to the nonzero couplings on the square links.}
\label{dbsq05}
\end{figure}

The phase diagrams of the system close to the square 
lattice limit and to the limit of two inter-penetrating square lattice
are shown in Fig.\ref{sq05} and Fig.\ref{dbsq05} respectively. 
They are very similar to the phase diagrams in their corresponding 
extreme limits except for the shifting of the phase boundaries.
It is clear from our phase diagrams that the ring exchange coupling
suppresses the homogeneous orbital antiferromagnetic state and favors 
the dimerized and/or inhomogeneous orbital antiferromagnetic state (box 
phase).

\section{Conclusion}
\label{sec:conclude}

In this paper, we studied the $t$-$J$ model with four-particle exchange 
couplings given by Eq.1 on an anisotropic triangular lattice as shown
in Fig.1. Even though the usual ring exchange limit corresponds to
$K=\lambda$ in Eq.1, we consider a more general model with $K\not=\lambda$
in order to get further insights. The large-$N$ mean field theory via
the $SU(N)$ generalization of the model is used to construct various
phase diagrams. The mean field ground states in the $N \rightarrow \infty$
limit can be classified using the link variables $\chi_{{\bf i}{\bf j}}$
given by Eq.9. The relevant link variables are shown in Fig.2. 

We are mostly interested in the effect of the four-particle exchange
couplings on the orbital antiferromagnetic state (staggered flux phase)
that exists in some parts of the large parameter space of the model.
It is found that it is useful to first consider two extreme limits;
the square lattice limit (all the parameters describing the
processes between the next-nearest-neighbor sites are zero) and the limit 
of inter-penetrating square lattice (all the parameters describing the
processes between the nearest-neighbor sites are zero).
This is mainly because, even in the ring exchange limit $K=\lambda$, the 
staggered flux phase exists as a stable ground states near these two limits; 
some small nonzero next-nearest-neighbor (nearest-neighbor) hopping and 
Heisenberg exchange couplings about the square lattice limit 
(the inter-penetrating square lattice limit). We studied these two extreme
cases as well as the conventional ring exchange limit at half-filling and
a finite hole concentration. 

Our results can be summarized as follows.
In the half-filled case ($\delta=0$), Fig.6 shows the comparison
of saddle-point energies of various phases in the square lattice limit.
In this case, the staggered flux phase is never a ground state
and either the fully dimerized phase or the box phase is the ground
state. One can see from Fig.6 that the difference between 
the saddle-point energies of the staggered flux phase and the ground state 
(dimerized/box phase) becomes larger for large $|K|$ and/or $|\lambda|$.    
At a finite hole concentration ($\delta=0.05$), the phase diagrams in the 
square lattice and the inter-penetrating square lattice limits are given 
by Fig.10 and Fig.15. Note that the finite hopping amplitudes can stabilize
the staggered flux phase that has circulating orbital currents (orbital
antiferromagnetic state) while the extremely large hopping amplitude 
eventually leads to the uniform phase. It is clear from these phase 
diagrams that the four-particle exchange processes favor the dimerized
(these are `partially' dimerized) phases or the inhomogeneous orbital 
antiferromagnetic state (box phase) over the homogeneous orbital 
antiferromagnetic state (staggered flux phase). 
This trend persists even in the conventional ring exchange limit $K=\lambda$
as can be seen from Fig.16 and Fig.17. Thus we conclude that the ring exchange 
or the four-particle exchange in general suppresses the staggered flux 
phase (homogeneous orbital antiferromagnetism), and instead favors the 
dimerized phases and/or the inhomogeneous orbital antiferromagnetic state 
depending on the value of $K$. 

We note that a recent quantum Monte Carlo study \cite{sandvik} 
of a quantum XY model with a ring exchange process discovered 
a striped bond-plaquette ordered state at intermediate values
of the ratio between the four-spin ring exchange and the nearest-neighbor
Heisenberg exchange couplings.
These results seem to be qualitatively consistent with 
our findings in the $t$-$J$-$K$ model studied here even though there are obvious 
differences in two models. The existence of more exotic phases \cite{balents1,senthil} 
like fractionalized phases in different parameter regimes of the $t$-$J$-$K$ model 
would be an interesting subject of future study.

\acknowledgements
This work was supported by the NSERC of Canada, Canada Research Chair
program of the NSERC, Canadian Institute for Advanced Research 
(H.Y.K. and Y.B.K.), and Alfred P. Sloan Fellowship (Y.B.K.).

\end{document}